Article type: **Full Paper**

# Tuning the Electronic Properties of LAO/STO Interfaces by Irradiating LAO Surface with Low-Energy Cluster Ion Beams

*Karl Ridier,[1,2,†*] Damien Aureau,[2] Bruno Bérini,[1] Yves Dumont,[1] Niels Keller,[1] Jackie Vigneron,[2] Arnaud Etcheberry,[2] Arnaud Fouchet[1,‡*]*

[1]Groupe d'Étude de la Matière Condensée (UMR 8635), Université de Versailles Saint-Quentin-en-Yvelines – CNRS – Université Paris-Saclay, 45 Av. des États-Unis 78035 Versailles, France.

[2]Institut Lavoisier de Versailles (UMR 8180), Université de Versailles Saint-Quentin-en-Yvelines – CNRS – Université Paris-Saclay, 45 Av. des États-Unis 78035 Versailles, France.

*Corresponding authors: arnaud.fouchet@ensicaen.fr, karl.ridier@hotmail.fr

[†] Present address: Laboratoire de Chimie de Coordination, CNRS UPR–8241, 205 route de Narbonne, F–31077 Toulouse, France.

[‡] Present address: Laboratoire de Cristallographie et Sciences des Matériaux (UMR 6508), Normandie Université, ENSICAEN (Ecole Nationale Supérieure d'Ingénieurs de Caen), UNICAEN (Université de Caen), CNRS, 6 Bd. Maréchal Juin, F–14050 Caen, France.

# Abstract

We have investigated the effects of low-energy ion beam irradiations using argon clusters on the chemical and electronic properties of LaAlO$_3$/SrTiO$_3$ (LAO/STO) heterointerfaces by combining X-ray photoelectron spectroscopy (XPS) and electrical transport measurements. Due to its unique features, we show that a short-time cluster ion irradiation of the LAO surface induces indirect modifications in the chemical properties of the buried STO substrate, with (1) a lowering of Ti atoms oxidation states (from Ti$^{4+}$ to Ti$^{3+}$ and Ti$^{2+}$) correlated to the formation of oxygen vacancies at the LAO surface and (2) the creation of new surface states for Sr atoms. Contrary to what is observed by using higher energy ion beam techniques, this leads to an increase of the electrical conductivity at the LAO/STO interface. Our XPS data clearly reveal the existence of dynamical processes on the titanium and strontium atoms, which compete with the effect of the cluster ion beam irradiation. These relaxation effects are in part attributed to the diffusion of the ion-induced oxygen vacancies in the entire heterostructure, since an increase of the interfacial metallicity is also evidenced far from the irradiated area. These results demonstrate that a local perturbation of the LAO surface can induce new properties at the interface and in the entire heterostructure. This study highlights the possibility of tuning the electronic properties of LAO/STO interfaces by surface engineering, confirming experimentally the intimate connection between LAO surface chemistry and electronic properties of LAO/STO interfaces.

# 1. Introduction

The quasi-2-dimensional electron gas (q-2-DEG) formed at the interface between the two band insulators LaAlO$_3$ (LAO) and SrTiO$_3$ (STO)[1] is one of the most fascinating systems in the field of oxide electronics. LAO/STO heterostructure is the subject of intensive research due to its multifunctional properties which open avenues for both fundamental and applied perspectives. Among its most interesting properties, the q-2-DEG is known to appear only when at least 3.5 units cells (u.c.) of LAO are epitaxially deposited on TiO$_2$-terminated STO substrates[2] and this interface has shown to host ferromagnetism,[3–6] two-dimensional superconductivity,[7] and strong spin-orbit coupling.[8,9] Despite intense interest, many basic questions remain about the microscopic mechanisms that give rise to the q-2-DEG at the LAO/STO interface. Pure intrinsic electronic reconstruction due to polar discontinuity (known as the "polar catastrophe" model),[10] presence of oxygen vacancies in the STO substrate,[11–13] or interfacial cation mixing[10,14,15] have been considered as possible mechanisms to, at least, partially explain the observed electronic properties of LAO/STO interfaces. Recently, an alternative model based on first-principles calculations,[16] suggests that the various experimental observations (including critical thickness, carrier density, interface magnetism) originate from an intricate balance between surface oxygen vacancies in LAO and cation antisite defects thermodynamically induced by the polar discontinuity across the interface.

While these mechanisms and their relative importance are still under debate, the LAO/STO heterostructure is particularly promising for applications because of the intrinsic confinement of the electronic properties at the interface and the possibility to control the electrical conductivity at the local scale. For instance, a tunable metal-insulator ground state has been evidenced in LAO/STO by using electric field effect.[2,17,18] Moreover, several observations reveal that the LAO surface (surface states and defects) strongly influences the properties of the interfacial q-2-DEG. As an example, the electronic properties of LAO/STO interfaces have been tailored by conducting atomic force microscopy,[19,20] by deposition of surface adsorbates on the LAO film,[21,22] by protonation of the LAO surface,[23] or by creation of surface defects as predicted by DFT calculations of LAO/STO heterostructure[16] and other interfaces like NdAlO$_3$/SrTiO$_3$.[24] This is an intriguing feature that a surface modification can remotely influence the interfacial electronic properties. A better understanding of this general effect is crucial: (1) in a fundamental point of view to bring new insights in the origin of the q-2-DEG and its properties, and (2) for applications as sensor or electronic devices, since small surface modifications or defects can lead to a sizeable change in the interfacial electronic properties.

Ion beam irradiation is a promising way to modify the surface properties of materials. In recent years, ion beam exposure has emerged as a novel tool to tailor the structural, mechanical, electronic, and even magnetic properties of materials by inducing atomic defects in a controlled manner. In particular, ion irradiation of oxide materials has been used for both device fabrication and material modification.[25,26] In LAO/STO, ion beam irradiation processes have been essentially used to induce metal-to-insulator (MIT) transition at the interface by creating defects (amorphization, local disorder) and strains in the heterostructure, either by using 150–350 eV monoatomic argon ions,[27,28] low-Z (2-MeV hydrogen and 500-keV helium) ions[29] or 50-keV oxygen ions[30] (see Table SI-1 in part 1 of Supporting Information). These techniques have been advantageously used for nano-patterning channels in which the original properties of the q-2-DEG (electrical conductivity, superconductivity) are preserved, opening interesting prospects for the realization of mesoscopic devices. In these previous studies, the MIT transition is described as the consequence of the large penetration lengths and experienced ion implantation which cause strong charge carrier localizations via ion-induced defects and structural changes in the heterostructure.

In this work, we explore the appealing features of *argon cluster ion beams* (annotated $Ar_n^+$ with n = 1000–2000 at./cluster) to irradiate the upper surface of LAO/STO heterostructures. Compared to monoatomic ion beam sputtering techniques, the main feature is that the average energy per argon atom is considerably low (only few eV per atom) leading to small penetration lengths into the sputtered material. The cluster ion beam sputtering technique has already been used on organic and polymer materials,[31,32] and as a promising tool to prepare oxide surfaces[33] or to perform depth profiling with low amount of defects in perovskite oxides like $SrTiO_3$.[34] In particular, this sputtering process has proved to strongly limit ion implantation in oxide materials.

We have combined X-ray Photoelectron Spectroscopy (XPS) and electrical measurements to investigate the effects of ion-beam-induced surface modifications on both the chemical and physical (electrical) properties of LAO/STO heterointerfaces. Due to its unique features, we demonstrate that this low-energy sputtering technique can be advantageously used to tune the chemical and electronic properties of LAO/STO interfaces. Due to the effective attenuation length L of the photoemitted electrons (L ≈ 2.05 nm in LAO and STO at the used photon energy),[35] XPS gives valuable information about the physicochemical properties of the buried interface through the LAO film. We have studied three 4.7-nm-thick (~ 12–13 u.c.) heterostructures (hereafter denoted H1, H2 and H3), grown under slight different conditions to provide interfaces with different initial electronic properties (see the experimental section for details). This thickness was chosen as a good compromise to have thick enough LAO film to preserve the interface from direct ion beam irradiation, and thin enough to probe the entire heterostructure by XPS even before any ion exposure.

In the first part of the article, the XPS depth-profiling analysis of a LAO/STO heterostructure by the Ar cluster ion beam sputtering technique is performed up to the complete etching of the LAO film. These first results give some insights on the evolution of the chemical and physical properties of the LAO/STO system during the cluster ion etching. In the second part, we have investigated the effects of a short-time cluster ion irradiation on the electronic behavior of LAO/STO interfaces and the possibility to tune the electrical properties of the q-2-DEG is addressed. Importantly, the crucial role played by ion-induced oxygen vacancies at the LAO surface as well as the existence of dynamical (relaxation) processes are clearly demonstrated.

# 2. XPS depth-profiling analysis of a LAO/STO heterostructure: effects of the cluster ion exposure on the chemical properties

In this first part, we have studied the effects of the argon cluster ion sputtering technique in the XPS depth-profiling analysis of the H1 heterostructure (see details in the experimental section). In this preliminary experiment, a complete etching of the LAO film has been performed. The use of Ar clusters (~ 1000 at./cluster) with a mean energy of 8000 eV (*i.e.* ~ 8 eV per incoming Ar atom) is necessary to completely etch the LAO layer in a reasonable time. The total counting time to obtain all XPS spectra with a satisfying signal-to-noise ratio between each etching step is about 12 min.

## 2.1. General results

Figure 1 shows the time evolution of atomic percentages measured by XPS along the cluster ion beam irradiation. The progressive and complete vanishing of the La and Al elements are evidenced, while, in the same time, the XPS signal of Sr and Ti atoms is enhanced. As already reported,[33,34] no signal of argon is detected by XPS during the depth-profiling, showing that the cluster ion exposure does not generate measurable implantation into the heterostructure in the XPS detection limit (%at [Ar] < 0.1 %).

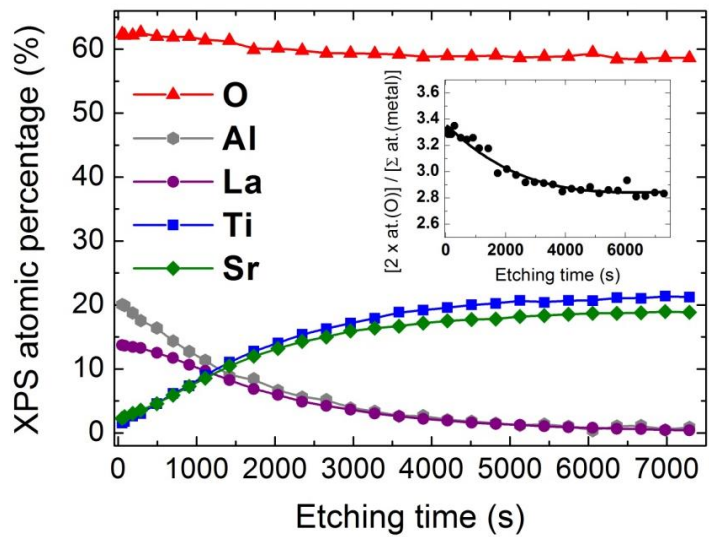

**Figure 1.** Time evolution of the XPS atomic percentage of the different detected elements along the argon cluster ion depth-profiling of the H1 heterostructure. Inset: time evolution of the ratio [2 × %at(O)] / [%at (Sr) + %at (Ti) + %at (La) + %at (Al)]. The solid line is a guide for the eye.

The time evolution of the LAO film thickness can be estimated along the depth-profile from the XPS atomic percentage of metal cations and the photoelectron effective attenuation lengths (EAL) in LAO and STO (see part 2.1 of Supporting Information). Although these calculations are approximate, we can estimate that the complete etching of the LAO film is reached after an irradiation of ~ 3000 s (*i.e.* with a sputter rate of about 0.95 Å/min in LAO). After 3000 s, a residual amount of La and Al is still detected likely because the interface is not perfectly abrupt and possibly destructured by the ion sputtering. Interestingly, a significant decrease of the atomic percentage of oxygen is observed from 62.2 % to 58.6 % throughout the depth profiling. This effect is better highlighted in the inset of Figure 1 which displays the evolution of the ratio [2 × %at (O)] / [%at (Sr) + %at (Ti) + %at (La) + %at (Al)] as a function of irradiation time. Although the sample H1 was annealed in oxygen after deposition, this ratio decreases by 15 % indicating the formation of a sizeable amount of oxygen vacancies in the heterostructure due to the cluster ion exposure.

The XPS spectra of the La 4d and Al 2p core-line peaks are presented in Supporting Information (Figure SI-1) at different times of the ion beam irradiation of H1. These data show the progressive decrease and the total vanishing of the La and Al peaks with sputtering time but no noticeable evolution is detected on the peak shape throughout the depth profiling. By comparison, as shown in Figure 2, strong modifications are evidenced on the XPS core-lines spectra of Ti $2p_{3/2}$ and Sr 3d. These peaks are displayed in Figures 2a and 2b, respectively, at different etching times. Their total intensity progressively increases with time, in the early stages of the irradiation, as the upper LAO film is progressively etched by the ion bombardment.

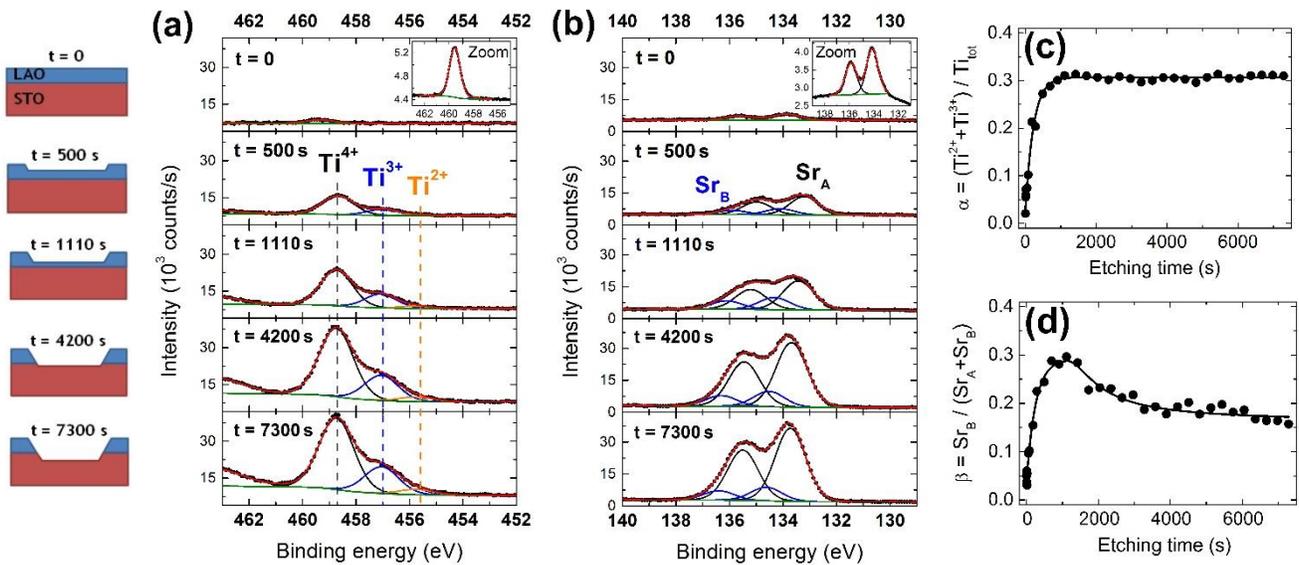

**Figure 2.** Influence of the cluster ion beam exposure on titanium and strontium atoms during the complete depth-profiling of the H1 heterostructure. **(a) and (b)**: XPS spectra of the Ti $2p_{3/2}$ and Sr 3d peaks, respectively, at different etching times. The sketches on the left schematically represent the depth of the ion etching at each step. The insets (at t = 0) show high-resolution (pass energy 10 eV) measurements of Ti $2p_{3/2}$ and Sr 3d spectra (with a counting time of ~ 2 h) performed prior to the beginning of the ion sputtering. The total counting time between each etching step was then fixed to 12 min. **(c) and (d)**: Evolution of the α = [$Ti^{3+}$ + $Ti^{2+}$] / [$Ti^{4+}$ + $Ti^{3+}$ + $Ti^{2+}$] and β = $Sr_B$ / [$Sr_A$ + $Sr_B$] ratios, respectively, as a function of etching time (solid lines are guides for the eye).

## 2.2. Titanium

As displayed in Figure 2a, the Ti $2p_{3/2}$ peak at t = 0 mainly consists of a $Ti^{4+}$ line while a very small component (< 2 %) of $Ti^{3+}$ can be eventually extracted from XPS data. As soon as the beginning of the cluster ion etching, we clearly observe the appearance of two low-binding energy components (at $\Delta E = 1.7 \pm 0.1$ eV and $3.0 \pm 0.1$ eV) associated to lower oxidation states ($Ti^{3+}$ and $Ti^{2+}$, respectively) of titanium atoms. Continuous lines in Figure 2a are the best fits assuming the same FWHM for the $Ti^{4+}$, $Ti^{3+}$ and $Ti^{2+}$ peaks. From the analysis of the fits, the time evolution of the ratio of the peak areas $\alpha = [Ti^{3+}+Ti^{2+}] / [Ti^{4+}+Ti^{3+}+Ti^{2+}]$ is represented in Figure 2c. This ratio grows rapidly to reach 30 % at t ≈ 1000 s and then remains stable until the end of the depth profiling. The change in the XPS Ti $2p_{3/2}$ spectra can also be correlated to the modifications observed in the valence band spectra (see Figure SI-2). In particular, the growth of a band of defect states occurs in the band gap region due to the cluster ion exposure. This feature has already been observed resulting from ion irradiation in STO.[34,36] These in-gap states signal the presence of 3d charge carriers near the interface in STO associated to the $Ti^{3+}$ and $Ti^{2+}$ ions.[37,38]

The lowering of oxidation states of titanium atoms is a well-known effect of ion beam exposures on STO substrates, especially using high-energetic monoatomic ions.[33,34,39,40] This effect is commonly attributed to the creation of oxygen vacancies in STO, which result in an insulator-to-metal transition and high-mobility surface conduction.[41–43] Indeed, each oxygen vacancy releases two electrons which can occupy the initially unoccupied Ti 3d band states,[36,44] resulting in $Ti^{3+}$ or even $Ti^{2+}$ low-binding energy components.

By way of comparison, we have carried out an ion beam irradiation using 8000 eV argon clusters on a bare STO substrate. This depth profile has been realized in the *same experimental conditions* than for the H1 LAO/STO heterostructure: same cluster size (1000 at./cluster), energy (8000 eV/cluster) and incoming current density (~ 15 µA/cm²). The results are presented in part 2.3 of Supporting Information. As shown in Figure SI-3, the cluster ion irradiation also induces lower oxidation states of Ti and thus generates oxygen vacancies into STO. Nevertheless, the α ratio reaches only 13.5 % in the stationary regime, *i.e.* a lower value than that obtained for the LAO/STO interface (30 %) (see discussion in SI). In LAO/STO, the large value of the ratio α and the occurrence of a significant $Ti^{2+}$ component signal the presence of a great amount of oxygen vacancies near the interface in the buried STO substrate. Interestingly, compared to the STO substrate, the creation of oxygen vacancies in the heterostructure seems to be significantly amplified by the presence of the upper LAO film.

## 2.3. Strontium

A careful investigation of the XPS Sr 3d core-line has also been conducted. As shown in Figure 2b, at t = 0 (see the inset), the Sr 3d peak is composed of one unique doublet arising from the spin-orbit coupling. The energy difference between the Sr $3d_{5/2}$ and Sr $3d_{3/2}$ components is around 1.76 eV and their intensity ratio is, as expected, equal to ~ 1.5. This observation signals the existence of a unique chemical environment for the Sr atoms in the vicinity of the as-deposited interface. In the same way as for the Ti $2p_{3/2}$ peak, the Sr 3d core-line is rapidly and strongly modified under the effect of the cluster ion beam exposure. XPS data clearly show the emergence of an additional high-binding energy doublet (hereafter denoted $Sr_B$) which progressively grows during the cluster ion sputtering. This $Sr_B$ component signals the appearance of a new chemical environment for the Sr atoms. Continuous lines in Figure 2b are the results of the best fits considering two doublets for the Sr 3d core-line. Fits were realized imposing the same FWHM for all components and an energy shift equal to 0.85 ± 0.05 eV between the two doublets.[39,40,45,46] The time evolution of the ratio of the doublet areas $\beta = Sr_B / [Sr_A + Sr_B]$ is presented in Figure 2d. This ratio increases up to 28 % after ~ 700 s of etching, reaches a small plateau between 700 s and 1500 s and finally decreases slowly to ~ 18 %.

Such a modification (growth of a high-binding energy doublet) of the Sr 3d core-line due to the ion beam irradiation has already been evidenced on STO substrates.[33,39,40] In STO, this $Sr_B$ doublet is associated to the existence of a surface component of Sr atoms which tend to migrate toward the outermost surface as a result of the energy supplied by the incoming Ar ions.[33,45,46] Interestingly, *the same effect occurs in LAO/STO* even though the STO substrate is covered by the LAO film. This has been evidenced by complementary low-energy $He^+$ ion scattering (LEIS) experiments. This technique is unique in its sensitivity to both structure and elemental composition of extreme surfaces and gives precise information on the chemical nature of the first atomic layer encountered by the helium ions projectiles. Figure 3 shows the LEIS spectra recorded on a pure STO substrate (black curve) and on LAO/STO at different etching times, during the first stages of the cluster ion exposure (t < 600 s, while the LAO film is not totally etched).

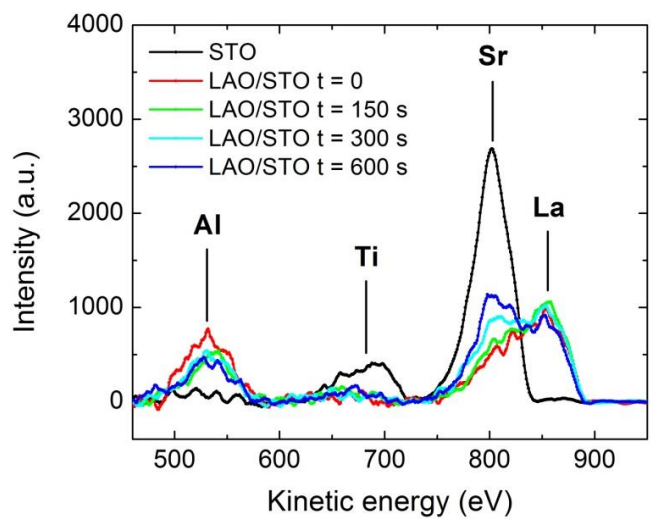

**Figure 3.** LEIS spectra (978-eV He⁺ ions, scattering angle of 130°) recorded on a STO surface (black curve) and a LAO/STO heterostructure at different times of the cluster ion beam exposure (2000 at./cluster, mean energy 4000 eV/cluster). Data have been normalized and corrected from the background.

The spectrum obtained on the pure STO substrate exhibits two main peaks around 680 and 800 eV corresponding to titanium and strontium atoms, respectively.[47] As for the LAO/STO heterostructure, the spectrum at t = 0 shows the existence of two peaks around 530 and 860 eV associated to aluminum and lanthanum atoms. Upon cluster ion exposure, a slight decrease of the Al intensity is evidenced as well as a gradual increase of the intensity around 800 eV, which is attributed to the presence of Sr atoms at the outermost LAO surface. No visible changes are observed in the Ti and La regions. These results strongly support the fact that the cluster ion exposure causes the migration of strontium atoms toward the outermost surface (through the LAO film), confirming the changes observed in the Sr 3d XPS spectra. Furthermore, by comparing the XPS results obtained (in the same experimental conditions) on a bare STO substrate (Figure SI-4), we can argue that the singular evolution of the ratio β (shown in Figure 2d) is also compatible with the formation of a Sr component at the LAO surface (see discussion in part 2.3 of Supporting Information).

## 2.4. Discussion

In this first part, we have investigated by XPS and LEIS the effects of an argon cluster ion beam exposure on the H1 heterostructure during the complete depth profiling of the LAO film. This low-energy sputtering technique has proved to be a "soft" etching technique which only alters the outermost surface of oxide materials (progressive etching with relatively low defect levels, low sputter rates and no argon implantation).[33,34] Despite the unique features of the cluster ion sputtering, strong modifications occur on the Ti and Sr atoms, as revealed by XPS data, while, interestingly, no visible change in the chemical environment of aluminum and lanthanum are detected during the ion exposure.

The most striking observation is that these modifications on Ti and Sr mainly occur in the first 700-900 s of etching, *i.e.* while the LAO film is not totally etched. At t = 900 s, the LAO thickness is, for instance, estimated to be ~ 3.0 nm. Therefore, chemical modifications on Ti and Sr atoms occur even though the STO substrate *does not undergo directly* the effects of the cluster ion irradiation. These observations strongly suggest that the incoming argon clusters cause defects at the LAO surface (while physical etching occurs) which in turn generate modifications in buried STO.

For titanium, the lowering of the mean oxidation state is clearly related to the formation of oxygen vacancies into the LAO/STO heterostructure. This is corroborated by the observation of a noticeable decrease of the oxygen content in the heterostructure during the etching process (see Figure 1). Moreover, as demonstrated by the comparative experiment performed on a bare STO substrate in the same irradiation conditions, these effects seem to be amplified by the presence of the upper LAO layer, presumably because the activation energy to create oxygen vacancies is lower in the LAO film deposited on STO than in the bare STO substrate. The most likely explanation is that a great amount of oxygen vacancies is created at the LAO surface of the heterostructure by the cluster ion exposure, generating a large concentration gradient. These oxygen vacancies are then transferred by diffusion into the STO, leading to a lowering of the oxidation states of buried Ti atoms. These diffusion phenomena will be more thoroughly evidenced and discussed in the second part of the article. As for strontium, XPS and LEIS data show the progressive migration of Sr atoms toward the outermost surface through the LAO film. The mechanism responsible for this effect is not completely understood. We can speculate that it may be related to a diffusion process, possibly stimulated by the selective sputtering of Al atoms in the LAO film by the cluster ion irradiation (*vide infra*).

These first observations support recent studies,[16] which predict that LAO/STO heterostructures are stabilized by the intricate balance of surface and interface defects. In the present case, the low-energy cluster ion beam irradiation creates new states at the LAO surface, which tend to destabilize the thermodynamic equilibrium of the heterostructure. Our XPS data show that the chemical properties of the interface (on the STO side) are then strongly affected to compensate the imbalance induced by the ion exposure. At this point, it appears intriguing to investigate the evolution of the electronic properties at the interface, in particular during the first steps of the ion-beam exposure when the chemical properties of STO are modified below the LAO film. This analysis is the purpose of the next part of this article.

# 3. Short-time argon cluster ion exposure on LAO/STO: monitoring of the interfacial conductivity

In this second part, we have investigated the evolution of the electronic properties of LAO/STO interfaces under the effect of a *short-time* argon cluster ion beam exposure. For that particular purpose, two 4.7-nm-thick LAO/STO samples (hereafter denoted H2 and H3) have been studied. Both interfaces were grown under identical conditions except that H2 was annealed for 1 h at T = 540 °C and P(O$_2$) = 0.2 mbar of pure oxygen immediately after deposition, whereas H3 was not annealed (see the experimental section for details). As a result, the two interfaces exhibit different initial electronic properties. On both samples, the LAO film was deposited on two distinct areas over the 10 × 4 mm$^2$ STO substrate, separated from each other by a 2-mm bare STO region. A schematic representation of H2 and H3 samples is given in Figure 4a. After deposition, AFM images confirmed the smoothness of the as-deposited surfaces and the presence of well-defined single-unit-cell high step terraces for both samples (see Figures SI-5a and SI-6a in Supporting Information).

The electrical transport properties of H2 and H3 are presented in Figures 4b and 4c. The temperature dependence of the sheet resistance is displayed in Figure SI-10. Initially, at room temperature, the sheet resistance of the "annealed" H2 sample ($R_S$ = 968 kΩ/□) is found to be two orders of magnitude larger than that of the "not annealed" sample H3 ($R_S$ = 17.5 kΩ/□) (dotted horizontal lines in Figure 4b). As displayed in Figure SI-10 (black curves), H3 presents a genuine metallic behavior characterized by an increase of the sheet resistance with temperature whereas the annealed sample H2 shows an insulating behavior. Carrier density measurements presented in Figure 4c confirm the observed differences between H2 and H3. Before any irradiation, the sheet carrier density of H3 at room temperature ($n_s$ = 7.0 × 10$^{13}$ cm$^{-2}$) is two orders of magnitude larger than that of H2 ($n_s$ = 2.8 × 10$^{11}$ cm$^{-2}$) (dotted lines in Figure 4c).

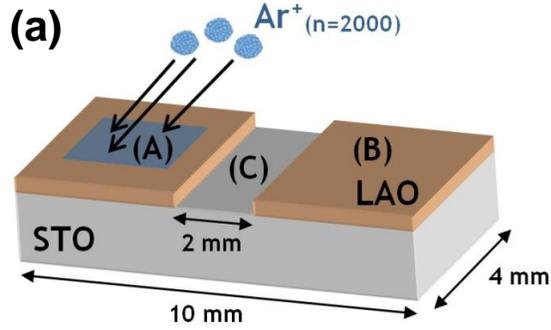
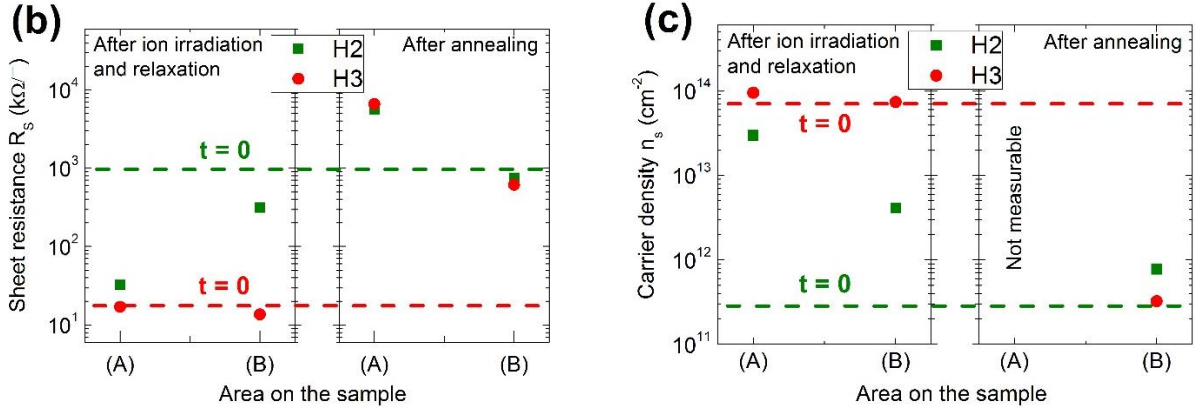

**Figure 4. (a)** Schematic representation of the H2 and H3 samples. For each sample, zone (A) is defined as the "ion-irradiated zone" and (B) as the "witness zone" on the other side of the 2-mm bare STO area, denoted (C). **(b) and (c)** Evolution of the sheet resistance and carrier density, respectively, measured at room temperature on areas (A) and (B) of heterostructures H2 and H3 after ion beam exposure (left panel) and after annealing (right panel). The dotted horizontal lines indicate the value of the sheet resistance and carrier density, respectively, before the ion beam irradiation (t = 0).

## 3.1. Cluster ion sputtering and relaxation processes

Ion sputtering has been realized by using 8000 eV argon clusters with a size (~ 2000 at./cluster) twice larger than for the depth-profiling of H1. The average energy per incoming argon atom was then ~ 4 eV. The two samples were irradiated on a square area, denoted (A), of approximately $3.25 \times 3.25$ mm$^2$ on one of the two deposited zone (see Figure 4a). Zone (B), located on the other side of the 2-mm bare STO area (C), can be defined as a "witness area". For both samples, the etching time was set to 580 s. At the end of the ion exposure, the thickness of the LAO film on the irradiated area (A) is estimated by XPS to be ~ 3.5 nm (~ 9 u.c.) – well above the critical thickness of 4 unit cells – ensuring that eventual modifications in the electronic properties cannot be imputed to physical etching of the LAO film below this critical thickness. XPS measurements were performed on the irradiated area (A) of H2 and H3 at different times of the ion beam exposure and during a waiting time of 800 min following the end of the sputtering. For this set of XPS measurements, the counting time between each etching step was increased to 68 min (instead of 12 min for H1) to improve the signal-to-noise ratio of the Ti 2p and Sr 3d spectra.

First, the question of argon implantation was investigated by XPS. As already mentioned for H1, a very tiny amount of implanted Ar is detected on the irradiated area (A) of both samples during the ion beam irradiation. Long counting times allow to extract a fraction of argon < 0.1 % of the total percentage of detected species (see Figure SI-7), confirming that no significant implantation occurs during the irradiation process. Ti 2p$_{3/2}$ and Sr 3d spectra are displayed in Figures SI-8 and SI-9, for H2 and H3, respectively, at different times of the experiment (at t = 0, at the end of the ion exposure, after the "relaxation" process and after final annealing). As previously observed for H1, the ion sputtering induces a lowering of the mean oxidation state of Ti (from Ti$^{4+}$ to Ti$^{3+}$) and the progressive growing of the surface component Sr$_B$ for strontium. The time evolution of the ratios of the peak areas: $\alpha$ = Ti$^{3+}$ / [Ti$^{4+}$ + Ti$^{3+}$] and $\beta$ = Sr$_B$ / [Sr$_A$ + Sr$_B$] is reported in Figures 5a and 5b, respectively.

Although H2 and H3 show similar trends during both etching and relaxation times, some differences can be noticed. As displayed in Figure 5a, at t = 0, $\alpha$ is found slightly larger for H3 ($\alpha \approx$ 3.6 %) than for H2 ($\alpha$ < 2 %), which is compatible with the initially insulating (conductive) character of the H2 (H3) interface. In both cases, $\alpha$ increases to reach a similar value (~ 8.2 %) after 580 s of etching. Then, a slight decrease of $\alpha$ is observed while waiting during 800 minutes in the XPS chamber. Finally, both interfaces reach a steady state in which the ratio of Ti atoms in lowered oxidation states is ~ 4.2 %. As shown in Figure 5b, a similar evolution occurs for strontium. The fraction $\beta$ = Sr$_B$ / [Sr$_A$ + Sr$_B$] grows continuously to reach ~ 11–13 % at the end of the etching. For both samples, $\beta$ becomes finally equal to 7–8 % in the steady state after relaxation processes.

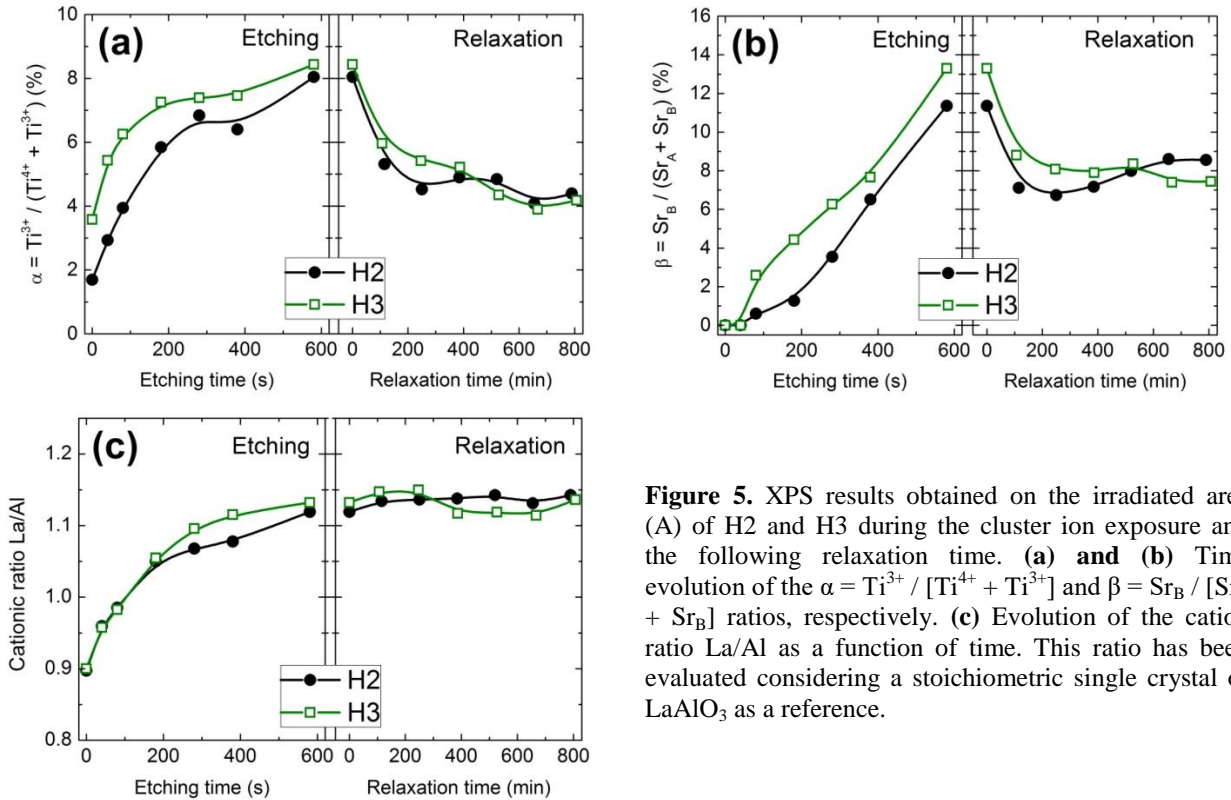

**Figure 5.** XPS results obtained on the irradiated area (A) of H2 and H3 during the cluster ion exposure and the following relaxation time. **(a) and (b)** Time evolution of the $\alpha$ = Ti$^{3+}$ / [Ti$^{4+}$ + Ti$^{3+}$] and $\beta$ = Sr$_B$ / [Sr$_A$ + Sr$_B$] ratios, respectively. **(c)** Evolution of the cation ratio La/Al as a function of time. This ratio has been evaluated considering a stoichiometric single crystal of LaAlO$_3$ as a reference.

The time evolution of α and β after the end of the cluster ion irradiation is a clear indication of the occurrence of dynamical processes on Ti and Sr atoms (on a time scale of hours) which compete with the effect of the ion exposure. Such relaxation effects have already been reported on titanium after $Ar^+$ monoatomic ion irradiation of STO substrates and were mainly attributed to the diffusion of ion-induced oxygen vacancies created near the surface.[34,42] Here, such a relaxation effect is evidenced for the first time on Ti and Sr atoms in LAO/STO. These dynamical processes are also the reason for the less pronounced chemical modifications on Ti and Sr atoms in H2 and H3 in comparison with H1. Indeed, as counting times were increased between each etching step (68 min instead of 12 min for H1), these relaxation phenomena compete with the ion beam irradiation, resulting in lower values for α and β.

Figure 5c shows that a modification of the cationic stoichiometry of the LAO film occurs during the ion sputtering. At t = 0, the La/Al ratio is 0.9 for both samples and it grows continuously up to 1.12–1.13. In other words, the LAO layer which was initially deficient in La (due to the PLD growth conditions: laser energy density, oxygen pressure, deposition temperature), shows La-rich compositions due to the ion irradiation. This suggests selective sputtering by the cluster ion beam and Al depletion within the LAO film, likely due to the important mass mismatch between La and Al ($m_{La} / m_{Al} \approx 5$). No relaxation process is observed in the cationic stoichiometry within the 800 minutes following the cluster ion beam irradiation.

After the cluster ion sputtering and the relaxation processes, we have investigated the electronic properties of both interfaces. As for H3, we find that the electrical properties are not noticeably modified on both irradiated (A) and "witness" (B) areas. Indeed, no significant change in both the sheet resistance (Figures 4b and SI-10) and the carrier density (Figure 4c) is evidenced. Conversely, the electrical properties of H2 are significantly modified since the sheet resistance on (A) is reduced by two orders of magnitude at room temperature (see Figure 4b and SI-10 for the temperature dependence of $R_S$). As a result, the irradiated area becomes almost as conductive as the non-annealed sample H3. As shown in Figure 4c, the carrier density is increased by two orders of magnitude on the irradiated area (A). These electrical measurements are consistent with our XPS data and an occupation of 3d levels on Ti atoms ($Ti^{3+}$) responsible for electrical conduction. Indeed, while the ratio α is not appreciably modified after ion exposure and relaxation for H3 (from 3.6 % at t = 0 to 4.2 %), it significantly increases for H2 (< 2 % at t = 0 to 4.2 %). An interesting finding is that the electrical resistance of H2 is also decreased on the "witness" area (B) but to a lesser extent (less than one order of magnitude at 300 K) than on the irradiated area (A), while the carrier density is increased by one order of magnitude. These observations demonstrate that the large amount of oxygen vacancies, created by the cluster ion beam irradiation on the area (A), diffuse throughout the STO substrate since effects of the ion exposure are detected as far as area (B).

Interestingly, in any cases the bare STO area (C) remains perfectly insulating, indicating that oxygen vacancies are undoubtedly stabilized and confined in the vicinity of the interface with LAO.

Finally, AFM images of the irradiated area for both samples show that the surface topography is strongly modified (see Figures SI-5b and SI-6b). The surface becomes granular and exhibits small mounds (~ 2 nm high) distributed over the sputtered surface.

### 3.2. After annealing

To confirm the primary role of oxygen vacancies and to check if electronic properties of the interfaces can be reversibly restored, the two samples H2 and H3 were then annealed (for 1 h at 530°C in an oxygen pressure of 0.2 mbar) in the same chamber used for the PLD. We observe that the witness area (B) of both samples retrieves the "normal" electrical properties of annealed interfaces: the electrical resistance (Figures 4b (right panel) and SI-10) and the carrier density (Figure 4c, right panel) measured on (B) are similar to the initial values obtained for the annealed H2 heterostructure before any irradiation. On the other hand, the irradiated areas (A) are found more insulating than at t = 0. For both samples, the sheet resistance is ~ 6 MΩ/□ at 300 K and its temperature dependence is characteristic of a strong localization regime (see Figure SI-10). No data for the carrier density are provided on the irradiated areas (A) after annealing, as the sheet resistance was too high to obtain reliable results.

Annealing has the effect of removing the oxygen vacancies present in the heterostructure. This has been confirmed by XPS measurements: after annealing, the $Ti^{3+}$ component of the Ti $2p_{3/2}$ peak is almost completely removed and the ratio α is found < 2 % (see Figures SI-8 and SI-9). Conversely, XPS spectra reveal that strontium is still modified on the ion-irradiated areas (A) of both samples. The ratio β is equal to 10.6 % and 13.2 % for H2 and H3, respectively, indicating that Sr atoms are still present at the LAO surface. In addition, AFM images of the irradiated area (A) (see Figure SI-6c) show that annealing has no visible effect on the surface topography which remains strongly modified. Likewise, the cationic stoichiometry of the LAO film is not changed by annealing and the irradiated LAO film is still La-rich.

### 4. General discussion and conclusion

In this second part, we have studied the evolution of the electronic properties of LAO/STO interfaces under the effect of a short-time argon cluster ion beam irradiation. This investigation has been carried out combining XPS and electrical transport measurements, which provide complementary information on both the chemistry and the physics of the heterointerface and possible correlations between them. Contrary to what was observed by using high-energy ion beam irradiation techniques, we find that the cluster ion beam exposure generally induces an increase of the electrical conductivity at the LAO/STO interface. This is undoubtedly correlated to the "soft" feature of this low-energy sputtering technique in perovskite oxide materials (low-energy per incoming Ar atom, progressive etching with relatively low defect levels, low sputter rate and negligible argon implantation).[33,34] Interestingly, even if the buried STO is *not directly* affected by the Ar clusters, XPS data revealed that the main effects are observed on Sr and Ti atoms near the interface. The lowering of the mean oxidation state of titanium and the apparition of a new chemical environment for strontium, attributed to a migration of Sr atoms toward the LAO surface, were clearly observed.

One of the first effect of the cluster ion exposure is to induce chemical modifications in the LAO film. Indeed, a clear increase of the La/Al ratio is measured on the irradiated areas. This reveals a selective sputtering of light Al atoms likely due to the large mass mismatch between La and Al ($m_{La} / m_{Al} \approx 5$). We can mention that such a selective sputtering by the cluster ion source has not been observed on STO substrates[34] because the mismatch between Sr and Ti is significantly smaller ($m_{Sr} / m_{Al} \approx 1.8$). Furthermore, it is worth to mention that the topography of the upper LAO surface is strongly modified by the cluster ion irradiation, as revealed by AFM.

In STO, the appearance of low oxidation states of Ti atoms is related to the formation of oxygen vacancies near the interface into the buried STO substrate. These oxygen vacancies are undoubtedly formed at the LAO surface before diffusing into STO. This assumption is supported by several arguments. Firstly, in similar irradiation conditions, we observe a significant larger fraction of $Ti^{3+}$ and $Ti^{2+}$ ions in the case of the LAO/STO interface than in a pure STO substrate, indicating that LAO plays a key role in the formation of oxygen vacancies. Secondly, the relaxation process (occurring on a time scale of hours) evidenced on the XPS Ti $2p_{3/2}$ core-line is consistent with a diffusion phenomenon of oxygen vacancies into STO and in the entire heterostructure. These results indicate that the LAO film acts as a magnifier in injecting oxygen vacancies into STO. Such a behavior has been confirmed by *ab-initio* calculations[48] which show that formation energies of oxygen vacancies in LAO are lower than in STO.

Buried strontium atoms are also remarkably affected by the cluster ion irradiation of the LAO surface. The formation of the high-binding energy component revealed by XPS is associated to the migration of Sr atoms toward the outermost surface (through the LAO film) due to the destabilization of the thermodynamic equilibrium of the heterostructure by the ion etching. Indeed, LEIS measurements revealed a noticeable enrichment of the upper surface with Sr atoms on the irradiated areas. It is worth to mention that migration of Sr toward the surface is a well-known property already observed in STO when some energy is provided to the system, either by high-temperature annealing[46] or by ion sputtering.[33,39,40] The mechanism at the origin of Sr migration in LAO/STO is not perfectly understood and further investigations should be done to explain this effect. We can speculate that the thermodynamic disturbance of the interface by the ion irradiation favors the migration of Sr into the LAO film in two different ways: La-Sr intermixing which is an expected phenomenon[49] or filling of ion-induced aluminum vacancies in LAO by Sr atoms. Otherwise, there is no apparent correlation between the modification rates of Ti and Sr. Indeed, as displayed in Figures 5a and 5b, the modification rate of Ti is much faster than that of Sr, in particular in the early stages of the etching, suggesting that the migration of Sr atoms is not directly related to the creation of oxygen vacancies.

All these chemical modifications detected by XPS are the cause of deep modifications in the electronic properties of the interface. Despite the change of Sr chemical environment and La/Al ratio, an increase of the electrical conductivity is observed at the LAO/STO interface after the ion beam irradiation. This effect is especially pronounced on the initially insulating interface H2 for which the sheet resistance is reduced by two orders of magnitude at room temperature. In the case of the non-annealed interface H3, the effect of the ion irradiation on the electrical properties is less pronounced because the interface is already saturated with oxygen vacancies.

Previous studies reported that only Al-rich LAO films result in conductive heterointerfaces.[50,51] Such outcomes are not verified in our case since the interfacial conductivity seems to be primarily driven by the formation of ion-induced oxygen vacancies. This point of view is supported by the fact that the removal of oxygen vacancies by post-annealing leads to a more insulating interface on the irradiated area. This effect cannot be imputed to physical etching of the LAO film below the 4-u.c. critical threshold for electrical conductivity because the estimated thickness (from XPS data) is ~ 3.6 nm (9–10 u.c.) after the cluster ion beam irradiation. Instead, several (combined) effects could explain this observation since the irradiated areas remain "chemically modified" after annealing. The modified cation stoichiometry (La/Al > 1),[50] the presence of strontium atoms at the surface or the modified surface topography could be many reasons to explain the increased resistivity. In addition, even if the cluster ion exposure does not act directly on STO, we cannot rule out the possible formation of defects and strains in STO in the vicinity of the interface that would cause carrier localization.

One other striking result of our study is the existence of dynamical processes, in concomitance with the cluster ion beam irradiation. These relaxation phenomena have been evidenced by XPS on both Ti and Sr peaks. For titanium, the relaxation effect is attributed to the diffusion of ion-beam-induced oxygen vacancies near the interface in the entire heterostructure. This effect has been confirmed by electrical transport measurements since an increase of the conductivity was also observed on the "witness area" (B) after the ion irradiation. The origin of the relaxation process observed on Sr atoms is more debatable, but it might be related to a diffusion process of Sr atoms in the LAO film. In any case, the fact that these dynamical processes occur on a time scale of hours clearly indicates that atomic relaxation phenomena should be considered instead of purely electronic relaxation processes.

Our findings have to be put in perspective compared to the results obtained by Aurino et al.[27,28] by using 150–350 eV argon monoatomic ions. In the first steps of the etching, the authors always observed an increase of the electrical resistance. It results from the creation of ion-beam-induced defects which eliminate the electrical conductivity faster than oxygen vacancies are created in the STO substrate. For longer etching times, the resistance starts to decrease again due to the formation of oxygen vacancies, but it stays at higher or similar values than before irradiation, certainly because a large amount of defects is finally created into the heterostructure through argon implantation. Here, due to the "soft" feature of the cluster ion sputtering, the regime in which the effects of the ion-beam-induced defects/strains (responsible for carrier localization) dominate over the creation of oxygen vacancies *is not observed*. Therefore, an increase of the electrical conductivity is always evidenced under cluster ion beam exposure.

In conclusion, this work gives new experimental insights into the physics of the LAO/STO interface and, in particular, into the influence of local surface perturbations on the electronic properties of the quasi-2-dimensional electron gas. Our results confirm the subtle connection between the LAO surface (via oxygen vacancies, cationic stoichiometry and defects) and the electronic properties of the q-2-DEG at the LAO/STO interface. Importantly, the primary role played by oxygen vacancies created at the LAO surface has been emphasized. The cluster ion beam irradiation technique can be seen as a mean to softly tune the electronic properties by injecting (locally) oxygen vacancies into the entire oxide heterostructure without inducing large amount of defects (implantation).

## 5. Experimental section

**Sample preparation:** The $LaAlO_3/SrTiO_3$ (001) samples were prepared by pulsed laser deposition (PLD) by ablating a single crystal LAO target onto $TiO_2$-terminated STO (001) substrates. The laser (KrF excimer laser, $\lambda = 248$ nm) energy density was set to 1.8 J/cm$^2$ with a repetition rate of 1 Hz. The substrates were heated to 740°C, and the oxygen partial pressure was set to $1.33\times10^{-5}$ mbar during deposition. Three 4.7-nm-thick (~ 12–13 u.c.) heterostructures (denoted H1, H2 and H3) have been grown.

Immediately after deposition, heterostructures H1 and H2 were annealed for 1 h at 540 °C and 0.2 mbar of pure oxygen whereas the sample H3 was not annealed.

**XPS measurements and cluster ion beam irradiation:** X-ray Photoelectron Spectroscopy (XPS) measurements were carried out using an ESCALAB 250 Xi spectrometer with a monochromatic Al-Kα X-ray source (hν = 1486.6 eV). The photoelectron effective attenuation lengths are very close in STO and LAO at this photon energy and can be taken equal to $L_{LAO} \approx L_{STO} \approx 2.05$ nm.[35] The detection was performed perpendicularly to the sample surface using a constant energy analyzer mode (pass energy 20 eV) and spectra were recorded with a 0.1 eV energy step size. Because of unavoidable charging effects, in particular in insulating samples, a low-energy electron flood gun was used for all samples, to carry out experiments in the same conditions. Because of resulting uncertainties in the absolute binding energy energies, all XPS spectra have been aligned considering the main O 1s core-line as the reference at 530.7 eV as reported in the literature.

Argon cluster ion beam irradiations were performed using the MAGCIS Dual Beam ion source. The average kinetic energy of each cluster (two sizes were used: ~ 1000 and ~ 2000 atoms per cluster) was fixed to 8000 eV. The incoming argon clusters reach the sample with an angle of 30° from the surface normal and the ion current density is fixed to ~ 15 μA/cm². LAO/STO heterostructures are irradiated on a square area of ~ 3.25 × 3.25 mm$^2$ while the size of the XPS analysis zone is 650 × 650 μm$^2$ in the center of this irradiated area. Quantification was performed based on the O 1s, C 1s, Ar 2p, Sr 3d, Ti 2p$_{3/2}$, La 4d and Al 2p peak areas after a Shirley type background subtraction using the Thermo Fisher Scientific Avantage© data system.

**Electrical characterization:** Sheet resistance measurements were carried out in a four-point configuration in the temperature range 2–300 K using a Quantum Design PPMS system. The interfaces were contacted by ultrasonic bonding with Al wire. The Al wire penetrates into the sample from the top surface and makes connection across the interface that enables us to probe the LAO/STO interfacial conductivity. Hall effect measurements were performed using the Van der Pauw method to extract the carrier density by applying a ± 9T magnetic field perpendicular to the interface plane. The resistivity data was anti-symmetrized according to the applied magnetic field in order to exclude the longitudinal $\rho_{xx}$ component.

**Atomic Force Microscopy (AFM):** Topography and roughness of the samples have been investigated by AFM (Brucker Dimension 3100) in tapping mode using commercial tips with 300 kHz resonant frequency and 40 N/m spring constant.

# Supporting Information

Supporting Information is available from the Wiley Online Library or from the author.

# Acknowledgements

This work was performed with the financial support of the programs OxyCAR II, Ile-de-France region DIM OxyMore for XPS measurements, and NOVATECH C'Nano Ile-de-France (Project No. IF-08-1453/R) for electrical measurements. K. R. is grateful to the Labex CHARMMMAT for post-doctoral grant.

# Keywords

# References

[1]    A. Ohtomo, H. Y. Hwang, *Nature* **2004**, *427*, 423.
[2]    S. Thiel, *Science* **2006**, *313*, 1942.
[3]    A. Brinkman, M. Huijben, M. van Zalk, J. Huijben, U. Zeitler, J. C. Maan, W. G. van der Wiel, G. Rijnders, D. H. A. Blank, H. Hilgenkamp, *Nat. Mater.* **2007**, *6*, 493.
[4]    Ariando, X. Wang, G. Baskaran, Z. Q. Liu, J. Huijben, J. B. Yi, A. Annadi, A. R. Barman, A. Rusydi, S. Dhar, Y. P. Feng, J. Ding, H. Hilgenkamp, T. Venkatesan, *Nat. Commun.* **2011**, *2*, 188.
[5]    J. A. Bert, B. Kalisky, C. Bell, M. Kim, Y. Hikita, H. Y. Hwang, K. A. Moler, *Nat. Phys.* **2011**, *7*, 767.
[6]    T. Taniuchi, Y. Motoyui, K. Morozumi, T. C. Rödel, F. Fortuna, A. F. Santander-Syro, S. Shin, *Nat. Commun.* **2016**, *7*, 11781.

[7] N. Reyren, S. Thiel, A. D. Caviglia, L. F. Kourkoutis, G. Hammerl, C. Richter, C. W. Schneider, T. Kopp, A.-S. Ruetschi, D. Jaccard, M. Gabay, D. A. Muller, J.-M. Triscone, J. Mannhart, *Science* **2007**, *317*, 1196.
[8] A. Caviglia, M. Gabay, S. Gariglio, N. Reyren, C. Cancellieri, J.-M. Triscone, *Phys Rev Lett* **2010**, *104*, 126803.
[9] M. Ben Shalom, M. Sachs, D. Rakhmilevitch, A. Palevski, Y. Dagan, *Phys. Rev. Lett.* **2010**, *104*.
[10] N. Nakagawa, H. Y. Hwang, D. A. Muller, *Nat. Mater.* **2006**, *5*, 204.
[11] W. Siemons, G. Koster, H. Yamamoto, W. A. Harrison, G. Lucovsky, T. H. Geballe, D. H. A. Blank, M. R. Beasley, *Phys. Rev. Lett.* **2007**, *98*.
[12] A. Kalabukhov, R. Gunnarsson, J. Börjesson, E. Olsson, T. Claeson, D. Winkler, *Phys. Rev. B* **2007**, *75*.
[13] G. Herranz, M. Basletić, M. Bibes, C. Carrétéro, E. Tafra, E. Jacquet, K. Bouzehouane, C. Deranlot, A. Hamzić, J.-M. Broto, A. Barthélémy, A. Fert, *Phys. Rev. Lett.* **2007**, *98*.
[14] P. R. Willmott, S. A. Pauli, R. Herger, C. M. Schlepütz, D. Martoccia, B. D. Patterson, B. Delley, R. Clarke, D. Kumah, C. Cionca, Y. Yacoby, *Phys. Rev. Lett.* **2007**, *99*.
[15] A. S. Kalabukhov, Y. A. Boikov, I. T. Serenkov, V. I. Sakharov, V. N. Popok, R. Gunnarsson, J. Börjesson, N. Ljustina, E. Olsson, D. Winkler, T. Claeson, *Phys. Rev. Lett.* **2009**, *103*.
[16] L. Yu, A. Zunger, *Nat. Commun.* **2014**, *5*, 5118.
[17] A. D. Caviglia, S. Gariglio, N. Reyren, D. Jaccard, T. Schneider, M. Gabay, S. Thiel, G. Hammerl, J. Mannhart, J.-M. Triscone, *Nature* **2008**, *456*, 624.
[18] D. F. Bogorin, P. Irvin, C. Cen, J. Levy, In *Multifunctional Oxide Heterostructures*; Tsymbal, E. Y.; Dagotto, E. R. A.; Eom, C.-B.; Ramesh, R., Eds.; Oxford University Press, 2012; pp. 364–388.
[19] C. Cen, S. Thiel, G. Hammerl, C. W. Schneider, K. E. Andersen, C. S. Hellberg, J. Mannhart, J. Levy, *Nat. Mater.* **2008**, *7*, 298.
[20] Y. Xie, C. Bell, T. Yajima, Y. Hikita, H. Y. Hwang, *Nano Lett.* **2010**, *10*, 2588.
[21] W. Dai, S. Adhikari, A. C. Garcia-Castro, A. H. Romero, H. Lee, J.-W. Lee, S. Ryu, C.-B. Eom, C. Cen, *Nano Lett.* **2016**.
[22] Y. Xie, Y. Hikita, C. Bell, H. Y. Hwang, *Nat. Commun.* **2011**, *2*, 494.
[23] K. A. Brown, S. He, D. J. Eichelsdoerfer, M. Huang, I. Levy, H. Lee, S. Ryu, P. Irvin, J. Mendez-Arroyo, C.-B. Eom, C. A. Mirkin, J. Levy, *Nat. Commun.* **2016**, *7*, 10681.
[24] X. Xiang, L. Qiao, H. Y. Xiao, F. Gao, X. T. Zu, S. Li, W. L. Zhou, *Sci. Rep.* **2014**, *4*.
[25] A. V. Krasheninnikov, K. Nordlund, *J. Appl. Phys.* **2010**, *107*, 071301.
[26] T. Wolf, N. Bergeal, J. Lesueur, C. J. Fourie, G. Faini, C. Ulysse, P. Febvre, *IEEE Trans. Appl. Supercond.* **2013**, *23*, 1101205.
[27] P. P. Aurino, A. Kalabukhov, N. Tuzla, E. Olsson, A. Klein, P. Erhart, Y. A. Boikov, I. T. Serenkov, V. I. Sakharov, T. Claeson, D. Winkler, *Phys. Rev. B* **2015**, *92*.
[28] P. Paolo Aurino, A. Kalabukhov, N. Tuzla, E. Olsson, T. Claeson, D. Winkler, *Appl. Phys. Lett.* **2013**, *102*, 201610.
[29] S. Mathew, A. Annadi, T. K. Chan, T. C. Asmara, D. Zhan, X. R. Wang, S. Azimi, Z. Shen, A. Rusydi, Ariando, M. B. H. Breese, T. Venkatesan, *ACS Nano* **2013**, *7*, 10572.
[30] S. Hurand, A. Jouan, C. Feuillet-Palma, G. Singh, E. Lesne, N. Reyren, A. Barthélémy, M. Bibes, J. E. Villegas, C. Ulysse, M. Pannetier-Lecoeur, M. Malnou, J. Lesueur, N. Bergeal, *Appl. Phys. Lett.* **2016**, *108*, 052602.
[31] J. L. S. Lee, S. Ninomiya, J. Matsuo, I. S. Gilmore, M. P. Seah, A. G. Shard, *Anal. Chem.* **2010**, *82*, 98.
[32] D. Rading, R. Moellers, H.-G. Cramer, E. Niehuis, *Surf. Interface Anal.* **2013**, *45*, 171.
[33] D. Aureau, K. Ridier, B. Bérini, Y. Dumont, N. Keller, J. Vigneron, M. Bouttemy, A. Etcheberry, A. Fouchet, *Thin Solid Films* **2016**, *601*, 89.
[34] K. Ridier, D. Aureau, B. Bérini, Y. Dumont, N. Keller, J. Vigneron, A. Etcheberry, A. Fouchet, *J. Phys. Chem. C* **2016**, *120*, 21358.
[35] M. P. Seah, *Surf. Interface Anal.* **2012**, *44*, 1353.
[36] V. E. Henrich, G. Dresselhaus, H. J. Zeiger, *Phys. Rev. B* **1978**, *17*, 4908.
[37] G. Drera, F. Banfi, F. F. Canova, P. Borghetti, L. Sangaletti, F. Bondino, E. Magnano, J. Huijben, M. Huijben, G. Rijnders, D. H. A. Blank, H. Hilgenkamp, A. Brinkman, *Appl. Phys. Lett.* **2011**, *98*, 052907.
[38] A. Koitzsch, J. Ocker, M. Knupfer, M. C. Dekker, K. Dörr, B. Büchner, P. Hoffmann, *Phys. Rev. B* **2011**, *84*.
[39] Y. Adachi, S. Kohiki, K. Wagatsuma, M. Oku, *Appl. Surf. Sci.* **1999**, *143*, 272.
[40] B. Psiuk, J. Szade, M. Pilch, K. Szot, *Vac. -Lond. THEN Oxf.- PERGAMON-* **2009**, S69–S72.
[41] D. Kan, T. Terashima, R. Kanda, A. Masuno, K. Tanaka, S. Chu, H. Kan, A. Ishizumi, Y. Kanemitsu, Y. Shimakawa, M. Takano, *Nat. Mater.* **2005**, *4*, 816.

[42]    M. Schultz, L. Klein, *Appl. Phys. Lett.* **2007**, *91*, 151104.
[43]    G. Herranz, O. Copie, A. Gentils, E. Tafra, M. Basletić, F. Fortuna, K. Bouzehouane, S. Fusil, é. Jacquet, C. Carrétéro, M. Bibes, A. Hamzić, A. Barthélémy, *J. Appl. Phys.* **2010**, *107*, 103704.
[44]    A. F. Santander-Syro, O. Copie, T. Kondo, F. Fortuna, S. Pailhès, R. Weht, X. G. Qiu, F. Bertran, A. Nicolaou, A. Taleb-Ibrahimi, P. Le Fèvre, G. Herranz, M. Bibes, N. Reyren, Y. Apertet, P. Lecoeur, A. Barthélémy, M. J. Rozenberg, *Nature* **2011**, *469*, 189.
[45]    P. A. van der Heide, Q. Jiang, Y. Kim, J. Rabalais, *Surf. Sci.* **2001**, *473*, 59.
[46]    K. Szot, W. Speier, U. Breuer, R. Meyer, J. Szade, R. Waser, *Surf. Sci.* **2000**, *460*, 112.
[47]    S. Gerhold, Z. Wang, M. Schmid, U. Diebold, *Surf. Sci.* **2014**, *621*, L1.
[48]    Y. Li, S. N. Phattalung, S. Limpijumnong, J. Kim, J. Yu, *Phys. Rev. B* **2011**, *84*.
[49]    A. Kalabukhov, Y. A. Boikov, I. T. Serenkov, V. I. Sakharov, J. Börjesson, N. Ljustina, E. Olsson, D. Winkler, T. Claeson, *EPL Europhys. Lett.* **2011**, *93*, 37001.
[50]    M. P. Warusawithana, C. Richter, J. A. Mundy, P. Roy, J. Ludwig, S. Paetel, T. Heeg, A. A. Pawlicki, L. F. Kourkoutis, M. Zheng, M. Lee, B. Mulcahy, W. Zander, Y. Zhu, J. Schubert, J. N. Eckstein, D. A. Muller, C. S. Hellberg, J. Mannhart, D. G. Schlom, *Nat. Commun.* **2013**, *4*, 2351.
[51]    E. Breckenfeld, N. Bronn, J. Karthik, A. R. Damodaran, S. Lee, N. Mason, L. W. Martin, *Phys. Rev. Lett.* **2013**, *110*.

# Supporting information of

Tuning the Electronic Properties of LAO/STO Interfaces by Irradiating LAO Surface with Low-Energy Cluster Ion Beams

*Karl Ridier,[1,2,†*] Damien Aureau,[2] Bruno Bérini,[1] Yves Dumont,[1] Niels Keller,[1] Jackie Vigneron,[2] Arnaud Etcheberry,[2] Arnaud Fouchet[1,‡*]*

[1]Groupe d'Étude de la Matière Condensée (UMR 8635), Université de Versailles Saint-Quentin-en-Yvelines – CNRS – Université Paris-Saclay, 45 Av. des États-Unis 78035 Versailles, France.

[2]Institut Lavoisier de Versailles (UMR 8180), Université de Versailles Saint-Quentin-en-Yvelines – CNRS – Université Paris-Saclay, 45 Av. des États-Unis 78035 Versailles, France.

*Corresponding authors: arnaud.fouchet@ensicaen.fr, karl.ridier@hotmail.fr

[†] Present address: Laboratoire de Chimie de Coordination, CNRS UPR–8241, 205 route de Narbonne, F–31077 Toulouse, France.

[‡] Present address: Laboratoire de Cristallographie et Sciences des Matériaux (UMR 6508), Normandie Université, ENSICAEN (Ecole Nationale Supérieure d'Ingénieurs de Caen), UNICAEN (Université de Caen), CNRS, 6 Bd. Maréchal Juin, F–14050 Caen, France.

# Part 1 of the article: Introduction

### 1.1) On the ion-beam irradiation experiments on LAO/STO heterostructures reported in the literature.

Ion-beam irradiation has been essentially used to induce metal-insulator (MIT) transition at the LAO/STO interface by creating defects (amorphization, local structural disorder) and strains in the heterostructure, either by using 150-eV monoatomic argon ions, low-Z (2-MeV hydrogen ions), or 50-keV oxygen ions. Table SI-1 summarizes the main results obtained during these previous ion-beam irradiation experiments performed on LAO/STO heterostructures.

| Ion beam irradiation | Ti reduction | Sr modification | Ion implantation | La/Al ratio | Observations | Mechanisms |
|---|---|---|---|---|---|---|
| 2-MeV proton beam[1] | ? | ? | ? | ? | Metal to insulator transition | - Creation of structural defects in STO<br>- Increased scattering centers at the interface |
| 50-keV oxygen ions beam[2] | ? | ? | yes | ? | Metal to insulator transition | - Structural changes and strain in STO due to a large penetration depth of the oxygen ions |
| 150-eV Ar$^+$ monoatomic ions beam[3,4] | no | no | yes | Increase | Metal to insulator transition | - Partial amorphization of the LAO film<br>- Ar implantation and local defects in STO<br>- No creation of oxygen vacancies (at least in the first steps of the irradiation) |
| 8000-eV argon cluster ions beam (~ 2000 at./cluster) [this work] | yes | yes | no | Increase | Insulator to metal transition | - Injection of oxygen vacancies<br>- No implantation / weak local structural disorder |

**Table SI-1.** Summary table of experimental ion beam irradiations performed on LAO/STO heterostructures.

# Part 2 of the article: XPS depth-profiling analysis of a LAO/STO heterostructure: effects of the cluster ion exposure on the chemical properties

## 2.1) Estimation of the LAO thickness from XPS measurements.

The thickness of the upper LAO film can be approximately estimated at each etching step from the intensity of the measured XPS peaks of cationic metals: Sr 3d, Ti $2p_{3/2}$, La 4d and Al 2p. From these intensities, the atomic percentage associated with each element detected by XPS can be deduced. According to the model developed by Seah,[5] the photoelectron effective attenuation length (EAL) at the used photon energy hv = 1486.6 eV is close in STO and LAO and can be taken equal to $L_{LAO} \approx L_{STO} \approx L \approx 2.05$ nm. Making the assumption of a perfectly abrupt interface between LAO and STO, the LAO thickness $e$ can be expressed in terms on the atomic percentages by the formula:

$$e = -L \ln\left(\frac{\%at_{Sr} + \%at_{Ti}}{\%at_{Sr} + \%at_{Ti} + \%at_{La} + \%at_{Al}}\right)$$

## 2.2) XPS depth-profiling of the H1 LAO/STO heterostructure.

In this part, we have studied the effects of the argon cluster ion sputtering technique (~ 1000 at./cluster, 8000 eV/cluster) in the XPS depth-profiling analysis of the H1 LAO/STO heterostructure. Figure SI-1 presents the XPS spectra of the La 4d and Al 2p core-line peaks at different times of the cluster ion beam irradiation of H1. The valence band XPS spectra are shown in Figure SI-2.

**Lanthanum and aluminum**

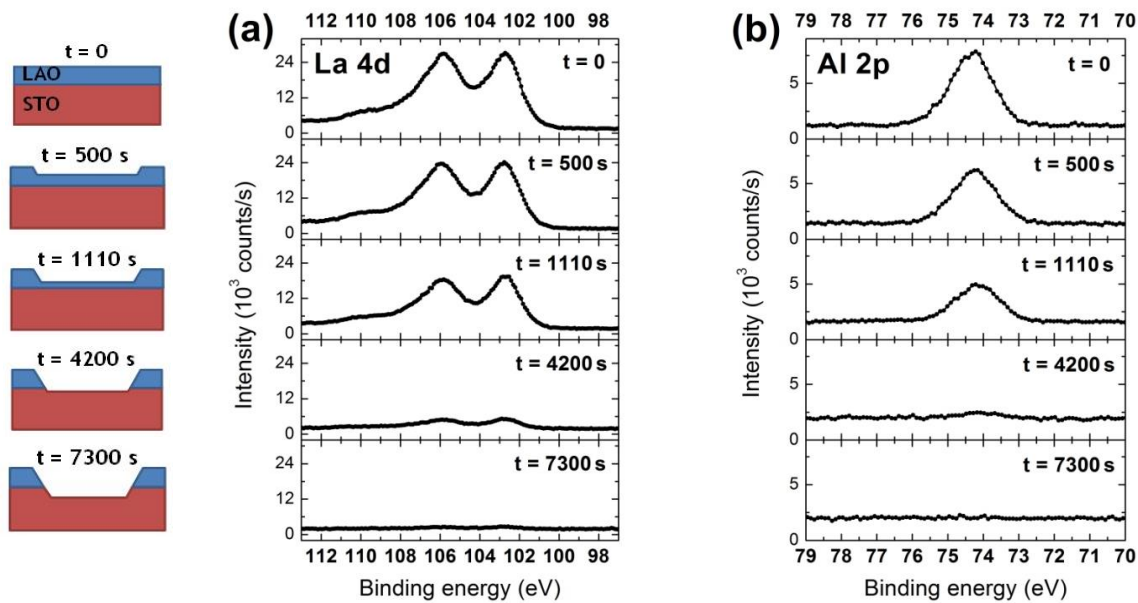

**Figure SI-1.** XPS spectra of the (a) La 4d peak and (b) Al 2p peak, respectively, at different times of the cluster ion sputtering of the H1 LAO/STO heterostructure. The sketches on the left schematically represent the depth of the ion etching at each step.

**Valence band**

As shown in Figure SI-2, the valence band is also strongly modified throughout the cluster ion sputtering of H1. The time evolution of the valence band is affected by the progressive disappearance of the LAO film with the decrease of the La 5p contribution and the enhancement of the Sr 4p component around 20 eV. The most important result is the progressive appearance of a band of defect states in the region of the band gap which signal the occupation of Ti 3d states. These in-gap states are clearly related to the lowering of oxidation states of titanium atoms ($Ti^{3+}$ and $Ti^{2+}$) evidenced by XPS.

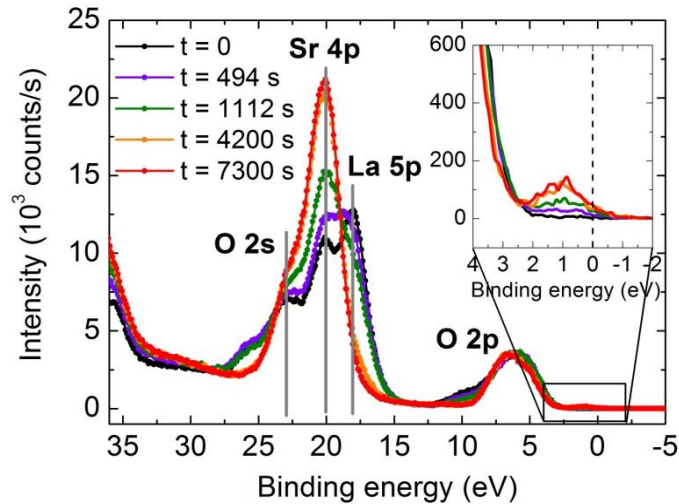

**Figure SI-2.** XPS spectra of the valence band region at different times of the argon cluster ion depth profiling (8000 eV, 1000 at./cluster) of the H1 LAO/STO heterostructure. The inset represents a zoom in the band gap region.

## 2.3) Comparison with a cluster ion irradiation (8000 eV/cluster, 1000 at./cluster) performed on a bare STO substrate.

To better understand the effects of the cluster ion beam exposure on the H1 LAO/STO heterostructure, we have carried out a depth profile of a bare STO substrate using the *same etching parameters*: clusters with a mean energy of 8000 eV, composed of ~ 1000 Ar atoms, reaching the sample with an angle of 30° with the same ion current density ~ 15 μA/cm$^2$. In addition, similar counting times and etching steps are chosen which is of great importance because of the occurrence of dynamical (relaxation) processes. Figures SI-3 and SI-4 display the XPS results obtained for titanium and strontium, respectively, on the bare STO substrate. These data are compared to those obtained on the H1 LAO/STO heterostructure presented in the main text of the article.

**Titanium**

XPS data (Figures SI-3a and SI-3b) show that the cluster ion beam exposure causes a lowering of the mean oxidation state of Ti atoms in the bare STO substrate. Nevertheless, by comparing with the spectra obtained on the H1 LAO/STO heterostructure after t = 2700 s of etching (see Figure SI-3b), we notice that the modification is much more pronounced in LAO/STO. For example, in no instance we observe a $Ti^{2+}$ component during the ion sputtering of the STO substrate while it is clearly evidenced in the case of LAO/STO. Similar FWHMs are obtained in both cases.

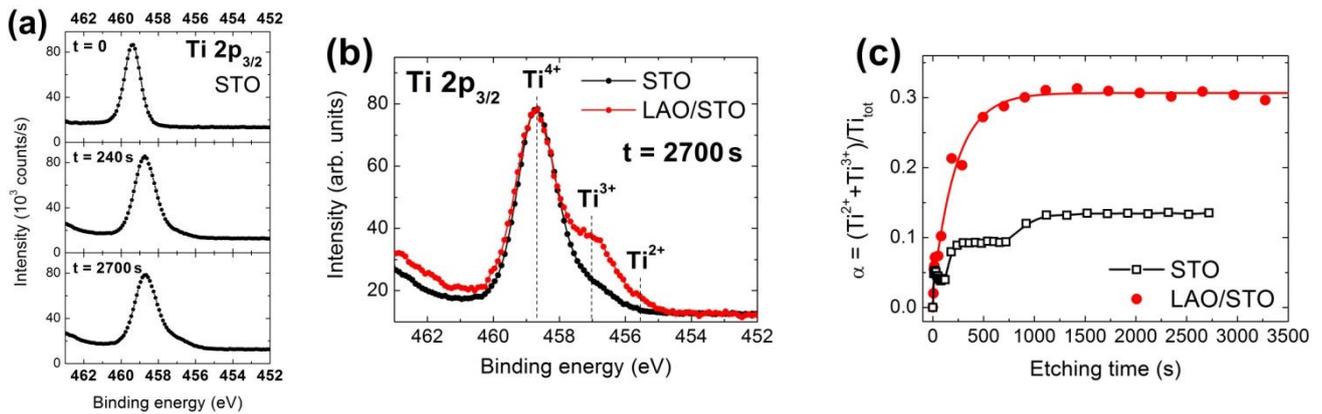

**Figure SI-3.** Comparative effects observed by XPS on titanium during argon cluster ion sputtering (8000 eV, 1000 at./cluster) of the H1 LAO/STO heterostructure and a bare STO substrate. **(a)** XPS spectra of the Ti $2p_{3/2}$ core-line at different etching times on STO. **(b)** Comparison of XPS spectra of the Ti $2p_{3/2}$ peak (normalized intensities) after 2700 s of etching. **(c)** Evolution of the ratio $\alpha = [Ti^{3+} + Ti^{2+}] / [Ti^{4+} + Ti^{3+} + Ti^{2+}]$ as a function of irradiation time.

Figure SI-3c shows the evolution of the ratio $\alpha = [Ti^{3+} + Ti^{2+}] / [Ti^{4+} + Ti^{3+} + Ti^{2+}]$ as a function of irradiation time. Interestingly, in STO, we observe that the stationary value of $\alpha$ depends on the duration of the etching steps: $\alpha$ is 9.5 % with etching steps of 60 s and 13.5 % with etching steps of 200 s. This clearly evidences the existence of dynamical processes (relaxation effects) competing with the ion sputtering. Such relaxation phenomena on titanium have already been reported after $Ar^+$ monoatomic ion irradiation of STO substrates and were mainly attributed to a diffusion process of ion-induced oxygen vacancies into the bulk material.[6,7] We have also put into evidence these dynamical effects in LAO/STO (see part 3 of the article). Finally, taking similar etching steps and counting times between each step, $\alpha$ reaches 13.5 % in STO and 30 % in LAO/STO. In other words, the fraction of Ti atoms in low oxidation states ($Ti^{3+}$ and $Ti^{2+}$) in LAO/STO is twice that of pure STO substrate.

**Strontium**

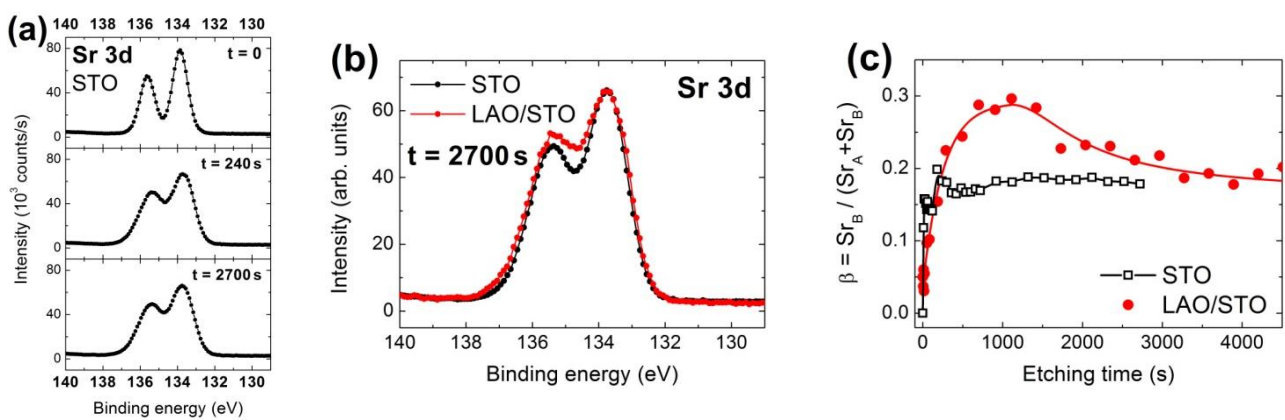

**Figure SI-4.** Comparative effects observed by XPS on strontium during argon cluster ion sputtering (8000 eV, 1000 at./cluster) of the H1 LAO/STO heterostructure and a bare STO substrate. **(a)** XPS spectra of the Sr 3d core-line at different times of the etching on STO. **(b)** Comparison of XPS spectra of the Sr 3d peak (normalized intensities) after 2700 s of etching. **(c)** Evolution of the ratio $\beta = Sr_B / [Sr_A + Sr_B]$ as a function of etching time.

As shown in Figure SI-4a, the XPS Sr 3d core-line peak experiences a clear modification under the effect of the cluster ion beam exposure in STO. Indeed, XPS data show the progressive growing of an additional high-binding energy doublet (denoted $Sr_B$) signaling the appearance of a new chemical environment for Sr atoms. This effect is a well-known feature observed in STO undergoing either Ar monoatomic or even cluster ion beam irradiation.[6,8–10] In STO, the high-binding energy doublet $Sr_B$ is usually associated to the existence of a surface component of Sr atoms which tend to migrate toward the outermost surface as a result of the energy supplied by the incoming Ar ions. This effect is, for instance, also observed after high temperature annealing of STO.[11–13]

As discussed in the main text of the article, similar modifications are evidenced on the XPS spectra of the Sr 3d peak during the cluster ion sputtering of the H1 heterostructure. The superposition of the XPS spectra in Figure SI-4b shows that at equal etching time (t = 2700 s), Sr atoms detected by XPS are apparently much strongly affected in LAO/STO than in the pure STO substrate. This is confirmed in Figure SI-4c which shows the evolution of the ratio $\beta = Sr_B / (Sr_A + Sr_B)$ as a function of irradiation time. In STO, the emergence of the $Sr_B$ doublet is faster (t < 150 s) and β reaches a constant value of ~ 18 %. By comparison, in LAO/STO, the increase of β is slower. It reaches a maximum up to 28 % before decreasing to tend to the value obtained in pure STO (~ 18 %). It is worth to note that β reaches this "STO stationary value" after t ~ 3500–4000 s, when the LAO film is completely etched.

This singular evolution of the ratio β in LAO/STO (going through a maximum then decreasing to the stable STO value) is compatible with the migration and the presence of Sr atoms at the LAO surface. Firstly, the increasing of β is slower than in STO because Sr atoms have to cross the LAO film to reach the upper surface. Secondly, β reaches a maximum value larger than in STO because the XPS signal of Sr atoms located at the surface $Sr_B$ is enhanced when the LAO film is present. In other words, it does not mean that the cluster ion beam exposure induces more Sr atoms at the surface than in STO. Finally, as the LAO layer is progressively removed, β decreases to attain the same value than in STO.

# Part 3 of the article: Short-time argon cluster ion exposure on LAO/STO: monitoring of the interfacial conductivity

### 3.1) AFM images of the irradiated areas.

Topography and roughness of the irradiated areas (A) has been investigated by atomic force microscopy (AFM) for both H2 (Figure SI-5) and H3 (Figure SI-6) heterostructures. After deposition, and before any ion irradiation, surfaces show well-defined terraces of 1-unit-cell high. After irradiation, the outermost surface is strongly modified: the terraces have disappeared and the surface becomes rough. In particular, we observe the appearance of small mounds (~ 2 nm high) uniformly distributed over the sputtered surface. This kind of topography seems to be a characteristic feature of cluster ion beam irradiated surfaces since it had been similarly evidenced after irradiation of pure STO substrates.[6] No further changes are observed after annealing.

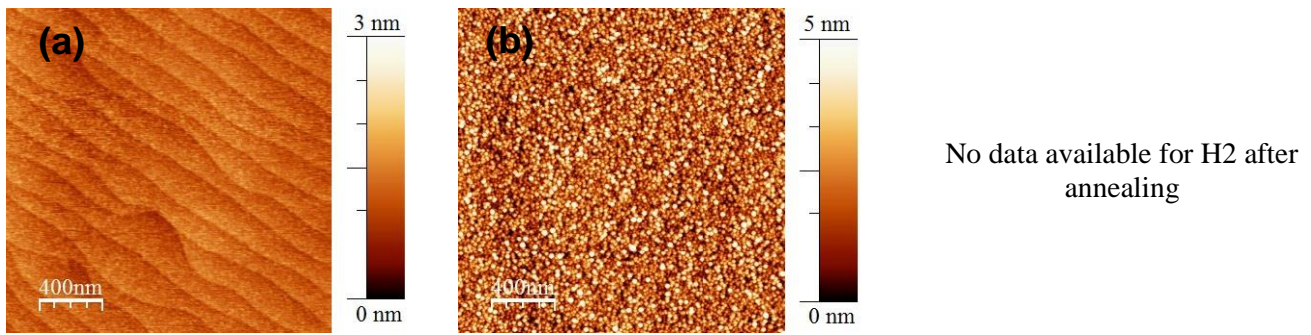

**Figure SI-5.** AFM images of the irradiated area (A) of sample H2 **(a)** after deposition and **(b)** after cluster ion beam exposure.

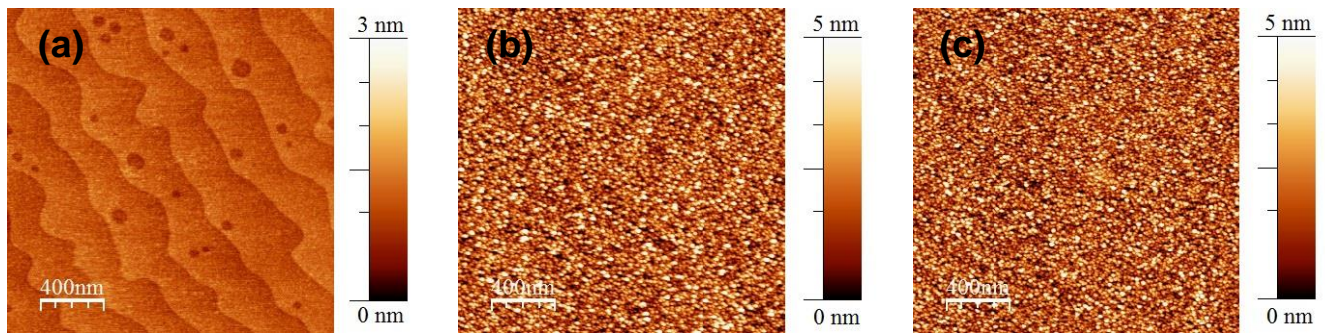

**Figure SI-6.** AFM images of the irradiated area (A) of sample H3 **(a)** after deposition, **(b)** after cluster ion beam exposure and **(c)** after annealing.

### 3.2) XPS analysis.

For this set of XPS measurements on H2 and H3, the total counting time was increased between each etching step (68 min instead of 12 min for H1) to improve the signal-to-noise ratio of the XPS Ti $2p_{3/2}$ and Sr 3d peaks. XPS measurements were also performed during a waiting time of 800 min following the end of the ion sputtering to detect possible changes in the XPS spectra. The evolution of the electrical properties of H2 and H3 was then investigated and, finally, further XPS measurements were made after an annealing of the two samples in pure oxygen.

**Argon implantation**

Implantation of argon was investigated by XPS during the cluster ion beam irradiation of H2 and H3. The XPS spectra recorded in the Ar 2p binding energy region, at t = 0 and at the end of the ion beam exposure, are shown in Figures SI-7a and SI-7b for H2 and H3, respectively. As displayed in Figure SI-7c, the atomic percentage of argon detected by XPS is very small and do not exceed 0.10 %. These observations confirm that argon implantation is negligible along the course of the cluster ion exposures.

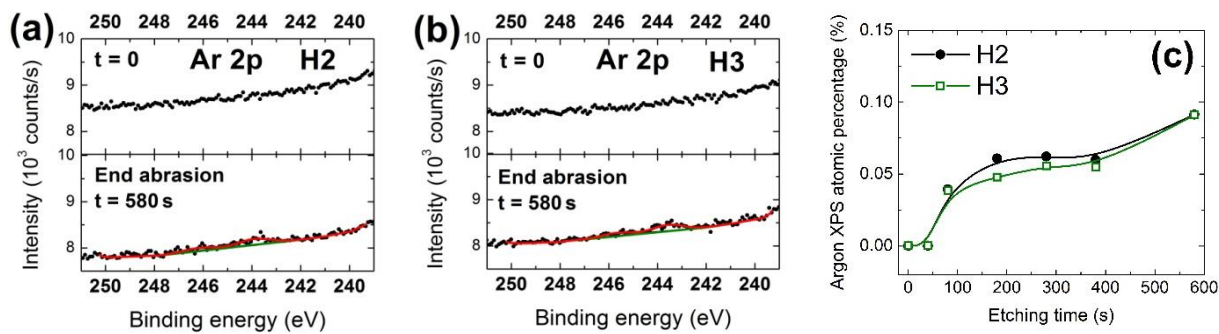

**Figure SI-7.** Argon XPS spectra. **(a) and (b)** XPS spectra of the Ar 2p binding energy region at t = 0 and t = 580 s (end of the ion exposure) for H2 and H3, respectively. **(c)** Time evolution of the argon XPS atomic percentage during the cluster ion beam irradiation of the two samples.

**Titanium and strontium**

Ti $2p_{3/2}$ and Sr 3d XPS spectra measured on the irradiated area (A) are presented in Figures SI-8 and SI-9 for H2 and H3 samples, respectively.

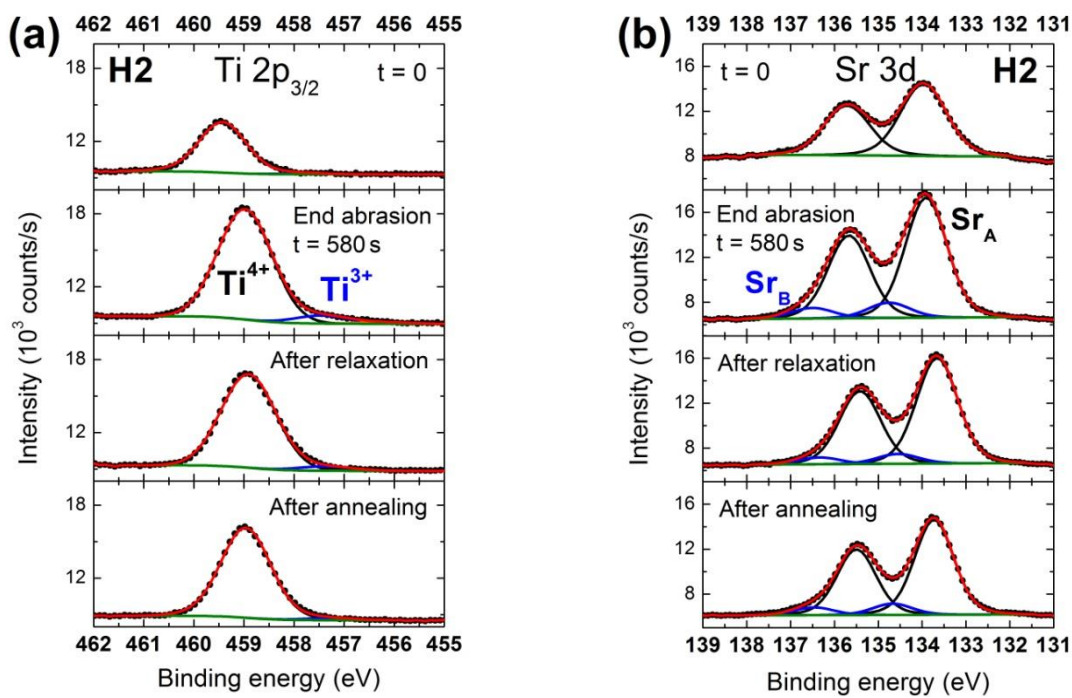

**Figure SI-8.** XPS spectra of the **(a)** Ti $2p_{3/2}$ peak and **(b)** Sr 3d peak recorded on the irradiated area (A) of H2 at different stages of the experiment (at t = 0, after the ion beam exposure, after the relaxation processes and after annealing).

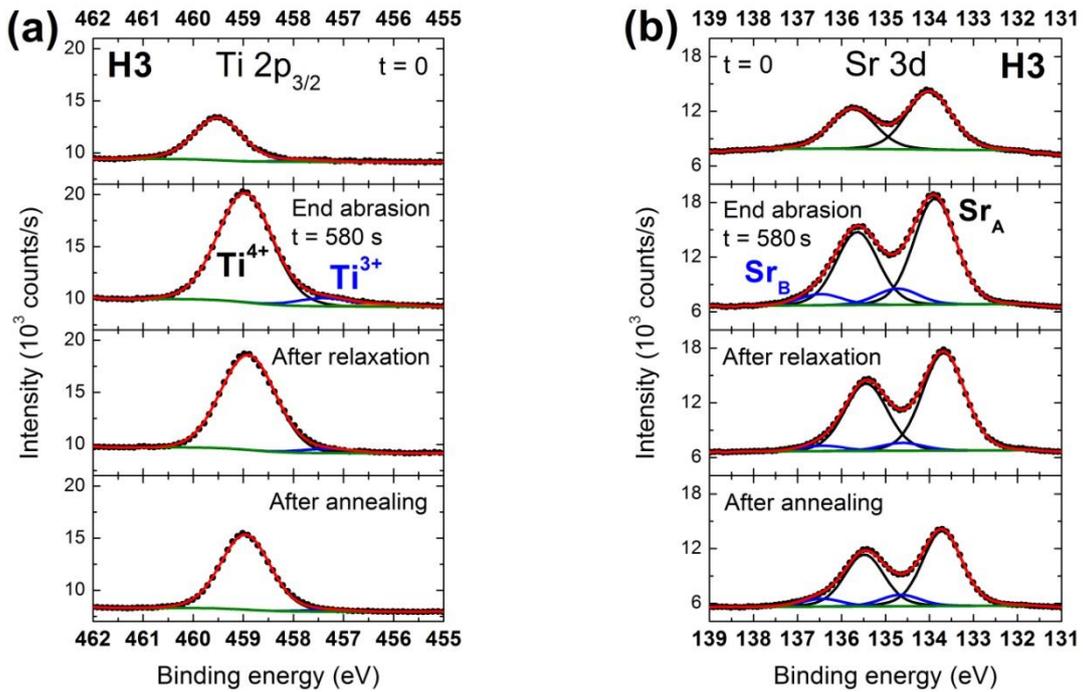

**Figure SI-9.** XPS spectra of the (**a**) Ti 2p$_{3/2}$ peak and (**b**) Sr 3d peak recorded on the irradiated area (A) of H3 at different stages of the experiment (at t = 0, after the ion beam exposure, after the relaxation processes and after annealing).

### 3.3) Electrical properties.

The thermal dependence of the sheet resistance has been measured in a four-point configuration on both areas (A) and (B) of samples H2 and H3 at t = 0 (initial properties before any ion irradiation), after the ion beam exposure and after annealing. The results are presented in Figure SI-10.

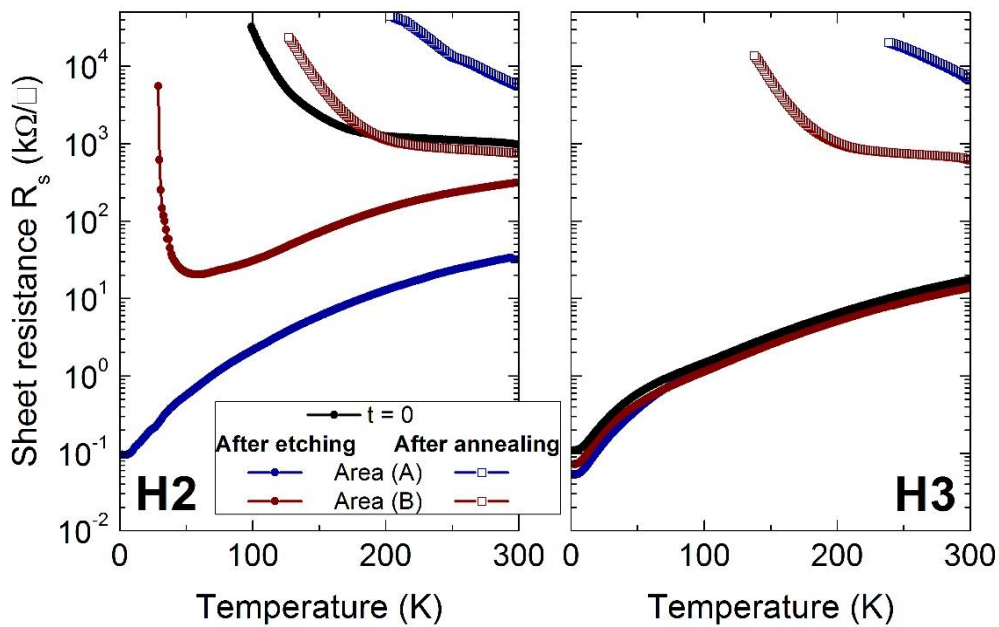

**Figure SI-10.** Temperature-dependent sheet resistance measured on areas (A) and (B) of H2 (left panel) and H3 (right panel) heterostructures, after the cluster ion beam exposure and after oxygen annealing. The black curve represents the sheet resistance before any ion beam irradiation (t = 0).

# References

[1]  S. Mathew, A. Annadi, T. K. Chan, T. C. Asmara, D. Zhan, X. R. Wang, S. Azimi, Z. Shen, A. Rusydi, Ariando, M. B. H. Breese, T. Venkatesan, *ACS Nano* **2013**, *7*, 10572.
[2]  S. Hurand, A. Jouan, C. Feuillet-Palma, G. Singh, E. Lesne, N. Reyren, A. Barthélémy, M. Bibes, J. E. Villegas, C. Ulysse, M. Pannetier-Lecoeur, M. Malnou, J. Lesueur, N. Bergeal, *Appl. Phys. Lett.* **2016**, *108*, 052602.
[3]  P. P. Aurino, A. Kalabukhov, N. Tuzla, E. Olsson, T. Claeson, D. Winkler, *Appl. Phys. Lett.* **2013**, *102*, 201610.
[4]  P. P. Aurino, A. Kalabukhov, N. Tuzla, E. Olsson, A. Klein, P. Erhart, Y. A. Boikov, I. T. Serenkov, V. I. Sakharov, T. Claeson, D. Winkler, *Phys. Rev. B* **2015**, *92*.
[5]  M. P. Seah, *Surf. Interface Anal.* **2012**, *44*, 1353.
[6]  K. Ridier, D. Aureau, B. Bérini, Y. Dumont, N. Keller, J. Vigneron, A. Etcheberry, A. Fouchet, *J. Phys. Chem. C* **2016**, *120*, 21358.
[7]  M. Schultz, L. Klein, *Appl. Phys. Lett.* **2007**, *91*, 151104.
[8]  B. Psiuk, J. Szade, M. Pilch, K. Szot, *Vacuum* **2009**, *83*, S69.
[9]  Y. Adachi, S. Kohiki, K. Wagatsuma, M. Oku, *Appl. Surf. Sci.* **1999**, *143*, 272.
[10] D. Aureau, K. Ridier, B. Bérini, Y. Dumont, N. Keller, J. Vigneron, M. Bouttemy, A. Etcheberry, A. Fouchet, *Thin Solid Films* **2016**, *601*, 89.
[11] P. A. W. Van der Heide, Q. D. Jiang, Y. S. Kim, J. W. Rabalais, *Surf. Sci.* **2001**, *473*, 59.
[12] K. Szot, W. Speier, U. Breuer, R. Meyer, J. Szade, R. Waser, *Surf. Sci.* **2000**, *460*, 112.
[13] H. Wei, L. Beuermann, J. Helmbold, G. Borchardt, V. Kempter, G. Lilienkamp, W. Maus-Friedrichs, *J. Eur. Ceram. Soc.* **2001**, *21*, 1677.